\documentclass[twocolumn]{aastex631}
\usepackage{CJK}
\usepackage{rotating}
\usepackage{afterpage}
\usepackage{changepage}
\usepackage{sidecap}

\newcommand{\squiggle}{SQuIGG$\vec{L}$E \,}
\newcommand{\squigglecomma}{SQuIGG$\vec{L}$E}

\newcommand{\pypeit}{\texttt{PypeIt}}

\defcitealias{bezansonNowYouSee2022}{BSS22}

\def\ntwo{[N\,{\sc ii}]}
\def\othree{[O\,{\sc iii}]}

\received{May 16, 2024}
\revised{Dec. 11, 2024}
\accepted{Jan. 27, 2025}

\submitjournal{ApJ}

\shortauthors{Zhu et al.}
\begin{document}
\begin{CJK*}{UTF8}{bsmi}
\title{\squiggle: Observational Evidence of Low Ongoing Star Formation Rates in Gas-Rich Post-Starburst Galaxies}

%\correspondingauthor{Mariska Kriek}
%\email{kriek@strw.leidenuniv.nl}

\author[0000-00  02-6768-8335]{Pengpei Zhu (朱芃佩)}
\affiliation{Leiden Observatory, Leiden University, P.O. Box 9513, 2300 RA Leiden, The Netherlands}
\affiliation{Cosmic Dawn Center (DAWN), Denmark}
\affiliation{DTU Space, Technical University of Denmark, Elektrovej 327, 2800 Kgs. Lyngby, Denmark}

\author[0000-0002-1714-1905]{Katherine A. Suess}
\affiliation{Department for Astrophysical \& Planetary Science, University of Colorado, Boulder, CO 80309, USA}
%\altaffiliation{NHFP Hubble Fellow}
%\affiliation{Kavli Institute for Particle Astrophysics and Cosmology and Department of Physics, Stanford University, Stanford, CA 94305, USA.}

\author[0000-0002-7613-9872]{Mariska Kriek}
\affiliation{Leiden Observatory, Leiden University, P.O. Box 9513, 2300 RA Leiden, The Netherlands}

\author[0000-0003-4075-7393]{David J. Setton}\thanks{Brinson Prize Fellow}
\affiliation{Department of Astrophysical Sciences, Princeton University, Princeton, NJ 08544, USA}

\author[0000-0001-5063-8254]{Rachel Bezanson}
\affiliation{Department of Physics and Astronomy and PITT PACC, University of Pittsburgh, Pittsburgh, PA 15260, USA}

\author[0000-0002-1759-6205]{Vincenzo Donofrio}
\affiliation{Department of Physics and Astronomy and George P. and Cynthia Woods Mitchell Institute for Fundamental Physics and Astronomy, Texas A\&M University, College Station, TX, USA}

\author[0000-0002-1109-1919]{Robert Feldmann}
\affiliation{Department of Astrophysics, University of Zurich, Winterthurerstrasse 190, Zurich CH-8057, Switzerland}

\author[0000-0003-4700-663X]{Andy D. Goulding}
\affiliation{Department of Astrophysical Sciences, Princeton University, Princeton, NJ 08544, USA}

\author[0000-0002-5612-3427]{Jenny E. Greene}
\affiliation{Department of Astrophysical Sciences, Princeton University, Princeton, NJ 08544, USA}

\author[0000-0002-7064-4309]{Desika Narayanan}
\affiliation{Department of Astronomy, University of Florida, 211 Bryant Space Science Center, Gainesville, FL 32611, USA}

\author[0000-0003-3256-5615]{Justin Spilker}
\affiliation{Department of Physics and Astronomy and George P. and Cynthia Woods Mitchell Institute for Fundamental Physics and Astronomy, Texas A\&M University, College Station, TX, USA}

%% Note that the \and command from previous versions of AASTeX is now
%% depreciated in this version as it is no longer necessary. AASTeX 
%% automatically takes care of all commas and "and"s between authors names.

%% AASTeX 6.31 has the new \collaboration and \nocollaboration commands to
%% provide the collaboration status of a group of authors. These commands 
%% can be used either before or after the list of corresponding authors. The
%% argument for \collaboration is the collaboration identifier. Authors are
%% encouraged to surround collaboration identifiers with ()s. The 
%% \nocollaboration command takes no argument and exists to indicate that
%% the nearby authors are not part of surrounding collaborations.

%% Mark off the abstract in the ``abstract'' environment. 
\begin{abstract}

ALMA observations have shown that {candidate} ``post-starburst" galaxies (PSBs) at z$\sim$0.6 can retain significant molecular gas reservoirs. These results would imply that -- unlike many model predictions -- {galaxies can} shut down their star formation before their cold gas reservoirs are depleted. However, these studies inferred star formation rates (SFRs) either from [O\,{\sc ii}] line fluxes or from spectral energy distribution modeling, and could have missed large dust-obscured contributions to the SFRs. In this study, we present Keck/NIRES observations of {13 massive ($\mathrm{M_*}\gtrsim \times 10^{11} \,\, \mathrm{M_\odot}$) PSBs}, which {allow us to estimate $\mathrm{H\alpha}$ SFRs} in these gas-rich post-starburst galaxies. We confirm the previously inferred low SFRs for the majority of the sample: 11/13 targets show clear $\mathrm{H\alpha}$ absorption, with minimal infilling indicating { dust-corrected} SFRs of $<4.1 \,\mathrm{M_\odot\, yr^{-1}}$. These SFRs are notably low given the large $\mathrm{H_2}$ reservoirs ($\sim 1-5 \times 10^{10} \,\, \mathrm{M_\odot}$) present in {5/13} of these galaxies, placing them significantly offset from star-forming galaxies {on the Kennicutt-Schmidt relation for star-forming galaxies}. { The [N\,{\sc ii}]/H$\alpha$ ratios of all 13 PSBs} imply contributions from non-star-forming ionization mechanisms {(e.g., AGN, shocks, or hot evolved stars)} to their $\mathrm{H\alpha}$ emission, suggesting that even these low ongoing SFRs may be overestimated. These low $\mathrm{H\alpha}$ SFRs{, dust-corrected using A$_v$ estimates from SED fitting,} confirm that these galaxies are very likely quiescent and, thus, that galaxies can quench before their cold gas reservoirs are fully depleted.

\end{abstract}

%% Keywords should appear after the \end{abstract} command. 
%% The AAS Journals now uses Unified Astronomy Thesaurus concepts:
%% https://astrothesaurus.org
%% You will be asked to selected these concepts during the submission process
%% but this old "keyword" functionality is maintained in case authors want
%% to include these concepts in their preprints.

%% From the front matter, we move on to the body of the paper.
%% Sections are demarcated by \section and \subsection, respectively.
%% Observe the use of the LaTeX \label
%% command after the \subsection to give a symbolic KEY to the
%% subsection for cross-referencing in a \ref command.
%% You can use LaTeX's \ref and \label commands to keep track of
%% cross-references to sections, equations, tables, and figures.
%% That way, if you change the order of any elements, LaTeX will
%% automatically renumber them.
%%
%% We recommend that authors also use the natbib \citep
%% and \citet commands to identify citations.  The citations are
%% tied to the reference list via symbolic KEYs. The KEY corresponds
%% to the KEY in the \bibitem in the reference list below. 

\section{Introduction} \label{sec:intro}

Galaxies can be generally divided into two types: blue, star-forming, gas-rich galaxies and red, quiescent, gas-poor galaxies. Yet the mechanisms that stop star formation and ``quench" galaxies remain unclear. For massive galaxies, the energy input from {active galactic nuclei (AGN)} is frequently invoked as a factor that disrupts the interstellar medium{, either by heating or removing it,} and thus suppresses star formation \citep[e.g.,][]{silkQuasarsGalaxyFormation1998,springelSimulationsFormationEvolution2005,rodriguezmonteroMergersStarburstsQuenching2019}. {These AGN may be triggered }by gas-rich major mergers \citep[e.g.,][]{wellonsFormationMassiveCompact2015}. {Alternative quenching} mechanisms have also been proposed, such as ``morphological quenching" where structural change prevents gas reservoirs from collapsing \citep[e.g.,][]{martigMORPHOLOGICALQUENCHINGSTAR2009}, or the incapability of the most massive systems to accrue and replenish their cold gas supplies \citep[e.g.,][]{feldmannArgoSimulationQuenching2015,daveMUFASAGalaxyStar2017}. Nearly all those models assume a tight connection between the cold $\mathrm{H_2}$ gas and ongoing star formation, {as observed in star-forming galaxies} at both large \citep[e.g.,][]{saintongeCOLDGASSIRAM2011,saintongeCOLDGASSIRAM2011a,tacconiPhibssMolecularGas2013,tacconiPHIBSSUnifiedScaling2018} and small scales \citep[e.g.,][]{kennicuttStarFormationGalaxies1998,schrubaMolecularStarFormation2011}. 

%Observational data indicates that quenching happens via two primary routes: a quick shift through the post-starburst phase or a more gradual transition through the green valley phase \citep[e.g.,][]{barroCANDELS3DHSTCompact2014,wildEvolutionPoststarburstGalaxies2016,suessDissectingSizeMassS1Mass2021}. 
{Observational results suggest that a significant portion of massive elliptical galaxies underwent star formation in early, short-lived episodes, highlighting the importance of swift quenching mechanisms \citep[e.g.,][]{thomasEpochsEarlyTypeGalaxy2005,kriekMassiveQuiescentPopulation2016, tacchellaFastSlowEarly2022,carnall24,degraaff24,beverage24}. Often, the end product of these rapid quenching processes are post-starburst (PSB) galaxies: these galaxies have gone through a recent episode of intensive star formation, which came to a halt within the last $\sim$1Gyr \citep[e.g.,][]{dresslerSpectroscopyGalaxiesDistant1983}.} From an observational standpoint, these galaxies' spectra typically display traits associated with A-type stars, such as strong Balmer breaks and deep Balmer absorption lines, while lacking evidence of instantaneous star formation {such as $\mathrm{H\alpha}$ emission lines. These spectral signatures indicate} that the burst concluded long enough ago for the most massive {O and B-type} stars to have perished, but recently enough for the {slightly} less massive, longer-lived {A-type} stars to still govern the optical spectrum \citep[e.g.,][]{zabludoffEnvironmentE+AGalaxies1996}.

Given that these galaxies are no longer {actively} forming stars, the expectation is that {post-starburst galaxies should have minimal remaining $\mathrm{H_2}$ reservoirs, similar to their older quiescent counterparts}. However, recent observations challenge this expectation both in the local universe and beyond: $\sim$half of local PSBs are found to contain substantial $\mathrm{H_2}$ reservoirs { with $f_\mathrm{H_2} \equiv M_\mathrm{H_2}/M_*$ reaching up to $\sim$50\% \citep[while the other $\sim$half are gas-poor; e.g.,][]{frenchDiscoveryLargeMolecular2015,rowlandsEvolutionColdInterstellar2015,alataloShockedPOststarburstGalaxy2016}}. %Yet, these local PSBs may not fully represent the processes that are responsible for quenching typical galaxies: most massive quiescent galaxies host very old stellar populations and quenched at much earlier times  %Since the most massive galaxies host the oldest stellar populations and therefore quenched earliest 
%\citep[e.g.,][]{mcdermidATLAS3DProjectXXX2015}. %, these remnants of the late-time quenching process at low redshift may not fully represent the processes that halted star formation in earlier epochs. 
 {A} small number of quiescent galaxies at $\mathrm{z > 0.1}$ have been probed for $\mathrm{H_2}$ using CO lines \citep{sargentDirectConstraintGas2015,spilkerMolecularGasContents2018,bezansonExtremelyLowMolecular2019,belliDiverseMolecularGas2021,zanellaLargeMolecularGas2023, parkRapidQuenchingGalaxies2023}, of which only 17/27 are recently-quenched post-starburst galaxies. %\citet{suessMassiveQuenchedGalaxies2017} reported two $z \sim 0.6$ PSBs that retain significant molecular gas reservoirs ($f_\mathrm{H_2} \sim$10\%-30\%), 
 Following the initial work by  \citet{suessMassiveQuenchedGalaxies2017}, \citet{bezansonNowYouSee2022} (hereafter BSS22) gathered a sample of 13 galaxies from the Studying QUenching in Intermediate-z Galaxies: Gas, angu$\vec{L}$ar momentum, and Evolution (\squigglecomma) program \citep{suessSQuIGGLEStudyingQuenching2022} and found that 6 galaxies {are gas-rich, with an average gas fraction} of $f_\mathrm{H_2} \sim$7\%-14\%. These studies all point towards molecular gas potentially being retained {after} quenching. {Most modern cosmological simulations assume a link between $\mathrm{H}_2$ and ongoing star formation \citep[e.g.,][]{crainEAGLESimulationsGalaxy2015, feldmannColoursStarFormation2017, daveMUFASAGalaxyStar2017, rodriguezmonteroMergersStarburstsQuenching2019}, { making it difficult for many theoretical models to explain how these large molecular gas reservoirs persist after star formation ceases.}}

{An alternative explanation for the observed gas reservoirs in post-starburst galaxies is that the galaxies are not actually quenched{ : these gas-rich galaxies may represent young, dust-obscured PSBs at early evolutionary stages \citep[e.g.][]{yesuf14,alataloEscapeAccretionStar2015,frenchClockingEvolutionPoststarburst2018,baronStarFormationMolecular2023}}.  }\citetalias{bezansonNowYouSee2022} uses star formation rate (SFR) estimates based on either [O\,{\sc ii}] line fluxes or stellar population fitting of the rest-optical SDSS spectra using a custom set of ``nonparametric" star formation histories (SFHs) \citep{leja19,johnsonStellarPopulationInference2021, suessRecoveringStarFormation2022}. %to retrieve estimates of the star formation rates (SFRs) of the sampled galaxies via \texttt{Prospector} \citep{johnsonBdJProspectorInitial2017,johnsonStellarPopulationInference2021} modeling.  
These measurements may miss highly dust-obscured star formation, meaning that the gas-rich nature of these objects may be explained by underestimated SFRs \citep[e.g.][]{belliDiverseMolecularGas2021, baronStarFormationMolecular2023}.

In this work, we present observations of the 13 \citetalias{bezansonNowYouSee2022} \squiggle PSBs with Keck II Telescope's Near-Infrared Echellette Spectrometer \citep[Keck/NIRES,][]{wilsonMassProducingEfficient2004}, focusing primarily on the region near $\mathrm{H\alpha}$ to measure the $\mathrm{H\alpha}$-based star formation rates (SFRs). {$\mathrm{H\alpha}$ provides a more direct and less dust-sensitive tracer of massive young stars than [O\,{\sc ii}] {and serves as a reliable indicator of instantaneous SFR \citep[$\sim$5-10Myr timescales; e.g.][]{sparre17,flores-velazquez21,tacchella22}, particularly for galaxies that are past their ``burst" phase \citep[e.g.,][]{genzelStudyGasstarFormation2010}. }These results imply that the $\mathrm{H\alpha}$ SFRs measured here will serve as a representative indicator of the current ongoing star formation in our sample, allowing us to investigate whether these galaxies currently have quiescent stellar populations as implied by their previously reported [O\,{\sc ii}] and SED SFRs.}

%a more robust SFR indicator than [O\,{\sc ii}], both because it is a more direct tracer of star-forming regions and because it is less affected by dust extinction.} Using the $\mathrm{H\alpha}$-SFRs, we have a better estimation of whether these galaxies indeed have quiescent stellar populations. %H$\beta$ emission is too weak to be reliably detected in the SDSS spectra of these galaxies, preventing us from performing a Balmer decrement correction; instead, we apply a dust correction to the $\mathrm{H\alpha}$ SFRs using the A$_v$ estimate from \texttt{FAST++} stellar population modeling.%, as H$\beta$ is not reliably detected in these galaxies to perform a Balmer decrement correction. Additionally, w
%We also examine the SDSS spectra of these galaxies, which allows us to measure the line ratios $\mathrm{[O\, III]\lambda5007/H\beta}$ and $\mathrm{[N\, II]\lambda6585/H\alpha}$ for a BPT diagram. %The SFRs estimated in this study were compared to those recovered by \citetalias{bezansonNowYouSee2022}, revealing a consistent correlation with very low SFR values, especially considering their molecular gas reservoirs.

We assume a $\Lambda$CDM cosmology with $\Omega_\Lambda$ = 0.7, $\Omega_m$ = 0.3, and $\mathrm{H}_0$ = 70 $\mathrm{km\, s^{-1}\, Mpc^{-1}}$, a \citet{chabrierGalacticStellarSubstellar2003} initial mass function, a \citet{calzettiDustContentOpacity2000} dust law, and adopt the vacuum wavelengths of emission lines from \citet{bylerNebularContinuumLine2017}.

\section{Data and Measurements} \label{sec: data}

\vspace{+0.1cm}

\subsection{The \squiggle Program} \label{subsec: squiggle}

The \squiggle program {consists of} a sample of 1318 recently-quenched massive PSBs at $0.50 < z < 0.94$ selected from the Sloan Digital Sky Survey (SDSS) Data Release 14 (DR14) spectroscopic database \citep{abolfathiFourteenthDataRelease2018} to have colors similar to A-type stars. {The sample selection and characterization are described in detail in \citet{suessSQuIGGLEStudyingQuenching2022}, but in brief: galaxies are selected within our target redshift range to have strong Balmer breaks (as probed by rest-frame $U-B$ color), blue slopes redward of the break (as probed by rest-frame $B-V$ color), and moderate- to high-S/N SDSS spectra. These cuts naturally produce a population with high $H\delta$ and low $D_n4000$, similar to other post-starburst selection methods \citep[e.g.][]{frenchDiscoveryLargeMolecular2015, wildEvolutionPoststarburstGalaxies2016}. \squiggle galaxies appear to have quenched recently and rapidly, and are characterized by specific SFRs (sSFRs) below $\sim10^{-11}\,\mathrm{yr}^{-1}$ (as inferred from their spectral energy distributions (SEDs)).} Spectral modeling shows that over 75\% of {the sample} formed at least a quarter of their total stellar mass during a recent burst, which concluded roughly 200 Myr before observation.

{Two sets of stellar population synthesis models have been published for these galaxies: \citet{settonSQuIGGEcLSurvey2020} presents models created using \texttt{FAST++}\footnote{\url{https://github.com/cschreib/fastpp}} \citep{kriekUltraDeepInfraredSpectrum2009}, and \citet{suessSQuIGGLEStudyingQuenching2022} presents \texttt{Prospector} \citep{johnsonBdJProspectorInitial2017,johnsonStellarPopulationInference2021} models. As described in detail in \citet{suessSQuIGGLEStudyingQuenching2022}, both sets of models are fit to the SDSS spectra and photometry as well as the Wide-Field Infrared Survey Explorer (WISE) 3.4 and 4.6$\mu$m photometry.  WISE 12 and 24$\mu$m points are not included in these fits: while the majority of \squiggle galaxies are not well-detected in these bands \citep{suessSQuIGGLEStudyingQuenching2022}, the galaxies that are detected have anomalously high mid-IR fluxes. Local studies of PSBs indicate that these high mid-IR fluxes are likely due to AGN, strong polycyclic aromatic hydrocarbon (PAH) features, warm dust, shocks, or post-AGB heating rather than obscured star formation \citep[e.g.,][]{alatalo17,smercina18,lee24}.}

{The 13 PSBs we study in this work are the subset of the full \squiggle sample that have published ALMA CO(2-1) observations. The ALMA data is presented in \citet{suessMassiveQuenchedGalaxies2017} and \citet{bezansonNowYouSee2022}; 6/13 galaxies are detected. As shown in \citet{bezansonNowYouSee2022}, the ALMA subset of galaxies is biased towards brighter $i$-band magnitudes but otherwise spans a relatively representative range of properties of the full \squiggle sample}. 
%presented in this work are a subset of the \squiggle sample that represents the stellar populations characteristic of the complete \squiggle sample. These galaxies all have ALMA CO(2-1) observations presented in \citetalias{bezansonNowYouSee2022}, with 6/13 are detected in CO.}

\subsection{The Keck/NIRES Observations and Reduction} \label{subsec: obs}

\vspace{-0.6cm}
\begin{deluxetable*}{cccccccccc}
    \tablecaption{Keck/NIRES \squiggle PSB Properties \label{tab:all measurements}}
    \tablecolumns{9}
    \tablewidth{0pt}
    \tablehead{
        \colhead{SDSS ID\tablenotemark{a}} & \colhead{\(z_{\mathrm{spec}}\)} & \colhead{$\log \frac{M_\mathrm{H_2}}{M_\odot}$\tablenotemark{b}} & \colhead{H\(\alpha\) flux\tablenotemark{c}} & \colhead{H\(\alpha\) EW} & \colhead{SFR} & \colhead{[O {\sc iii}]/H\(\beta\)} & \colhead{[N {\sc ii}]/H\(\alpha\)} & \colhead{\(A_{\mathrm{V, H_{II}}}\)\tablenotemark{d}} & \colhead{\(\sigma_v\)\tablenotemark{e}}\\
        & & & $10^{-16} \,\mathrm{erg/s/cm^2}$ & $\mathrm{\AA}$ & \(\mathrm{M_\odot}/\mathrm{yr}\) & & & mag & km/s
    }
    \startdata
    SDSS\_J2258$+$2313 & 0.706 & 10.72$\pm$0.02 & $3.08_{-0.34}^{+0.33}$ & $4.56_{-0.50}^{+0.49}$ & $7.11_{-1.89}^{+1.88}$ & $-$ & $0.49_{-0.12}^{+0.12}$ & $1.10\pm0.24$ & 122.7 \\
    SDSS\_J1109$-$0040 & 0.594 & 10.19$\pm$0.03 & $1.27_{-0.13}^{+0.13}$ & $2.44_{-0.25}^{+0.25}$ & $2.24_{-0.68}^{+0.68}$ & $<4.25$ & $1.35_{-0.15}^{+0.17}$ & $1.30\pm0.28$ & 150.6 \\
    SDSS\_J1007$+$2330 & 0.635 & $<9.53$ & $1.11_{-0.26}^{+0.22}$ & $1.90_{-0.44}^{+0.38}$ & $1.44_{-0.39}^{+0.36}$ & $-$ & $1.31_{-0.27}^{+0.32}$ & $0.67\pm0.14$ & 143.7 \\
    SDSS\_J2202$-$0033 & 0.592 & 9.81$\pm$0.06 & $1.06_{-0.18}^{+0.18}$ & $1.55_{-0.27}^{+0.26}$ & $1.51_{-0.35}^{+0.34}$ & $<4.11$ & $1.16_{-0.23}^{+0.29}$ & $0.69\pm0.15$ & 155.3 \\
    SDSS\_J0753$+$2403 & 0.565 & $<9.22$ & $1.02_{-0.12}^{+0.12}$ & $1.93_{-0.23}^{+0.23}$ & $1.97_{-0.73}^{+0.73}$ & $1.24_{-0.43}^{+0.94}$ & $1.56_{-0.19}^{+0.23}$ & $1.58\pm0.34$ & 118.0 \\
    SDSS\_J0046$-$0147 & 0.609 & $<9.39$ & $0.77_{-0.10}^{+0.10}$ & $1.35_{-0.17}^{+0.17}$ & $1.57_{-0.53}^{+0.53}$ & $-$ & $1.03_{-0.16}^{+0.19}$ & $1.41\pm0.30$ & 127.4 \\
    SDSS\_J1302$+$1043 & 0.592 & 10.19$\pm$0.05 & $0.68_{-0.22}^{+0.23}$ & $1.08_{-0.35}^{+0.37}$ & $0.55_{-0.18}^{+0.19}$ & $-$ & $1.61_{-0.48}^{+0.76}$ & $0.28\pm0.06$ & 162.2 \\
    SDSS\_J0027$+$0129 & 0.585 & $<9.38$ & $0.49_{-0.22}^{+0.22}$ & $0.74_{-0.33}^{+0.33}$ & $0.45_{-0.20}^{+0.21}$ & $<8.24$ & $3.10_{-1.02}^{+2.24}$ & $0.47\pm0.10$ & 120.3 \\
    SDSS\_J0233$+$0052 & 0.592 & $<9.35$ & $0.36_{-0.12}^{+0.11}$ & $0.62_{-0.20}^{+0.19}$ & $0.45_{-0.17}^{+0.16}$ & $-$ & $2.12_{-0.51}^{+0.85}$ & $0.84\pm0.18$ & 136.7 \\
    SDSS\_J1203$+$1807 & 0.595 & $<9.26$ & $0.27_{-0.13}^{+0.15}$ & $0.59_{-0.28}^{+0.32}$ & $0.37_{-0.19}^{+0.21}$ & $-$ & $2.04_{-0.80}^{+1.63}$ & $0.95\pm0.20$ & 148.3 \\
    SDSS\_J0912$+$1523 & 0.747 & 10.53$\pm$0.02 & $<1.10$ & $<1.72$ & $<4.28$ & $-$ & $<6.27$ & $1.60\pm0.34$ & 166.9 \\
    SDSS\_J1053$+$2342 & 0.637 & $<9.54$ & $<0.60$ & $<1.16$ & $<0.66$ & $-$ & $<7.15$ & $0.47\pm0.10$ & 113.3 \\
    SDSS\_J1448$+$1010\tablenotemark{f} & 0.646 & 10.29$\pm$0.03 & $23.06_{-0.81}^{+0.83}$ & $30.27_{-1.08}^{+1.09}$ & $31.06_{-4.72}^{+4.73}$ & $>5.67$ & $1.26_{-0.07}^{+0.07}$ & $0.67\pm0.14$ & 364.6 \\
    \enddata
    \tablenotetext{a}{ IDs match the publicly-available \squiggle catalog in \citet{suessSQuIGGLEStudyingQuenching2022}, which contains additional information such as precise coordinates.}
    \tablenotetext{b}{The CO(2-1) measured \(H_2\) gas mass from \citet{bezansonNowYouSee2022}, assuming $\alpha_\mathrm{CO} = 4.0$.}
    \tablenotetext{c}{Dust-corrected H$\alpha$ flux.}
    \tablenotetext{d}{ Assuming $A_{\mathrm{V,H_{\, II}}}$ / $A_{V,*}=1.86$ following \citealt{priceDIRECTMEASUREMENTSDUST2014}.}
    \tablenotetext{e}{The fitted line width on the target's original emission spectrum, can only serve as a representation of the line width.}
    \tablenotetext{f}{The SFR measurement of J1448+1010 is expected to be significantly overestimated given its AGN presence.}
    \tablecomments{Upper and lower limits for the undetected sources are \(2\sigma\).}
\end{deluxetable*}

{Thirteen \squiggle PSBs were observed with Keck/NIRES over four dates spanning 2018 to 2019. NIRES has simultaneous wavelength coverage from 0.94-2.45$\mu$m, a mean resolution of 2700, and a fixed longslit (0\farcs55 x 18\farcs). Total integration times for each object ranged from 45-65 minutes, with longer integrations for targets with fainter $i$-band magnitudes. All targets are sufficiently bright to perform target acquisition directly using the NIRES camera, without the use of blind offsets; centering of the object in the slit was checked periodically throughout each integration. Atmospheric seeing ranged from 0\farcs55 - 1\farcs1 depending on the date of observation. {Each target was observed using an ABC dither pattern with an offset of 2\farcs5 between A \& B and B \& C}, which improves the S/N of the reduced spectra compared to the more common ABBA pattern \citep[see the Appendix of][]{kriekMOSFIREDEEPEVOLUTION2015}. }Multiple frames of telluric stars were also collected for each observation date. {Tellurics were taken with a standard ABBA dither pattern to avoid detector persistence at the location of the ABC dithers for the primary targets.}
Table \ref{tab:all measurements} displays each target's name and measured physical properties.  %{ Each target was observed  } We employed a 2.5" A-B-C dither pattern for each target, which improves the S/N of the reduced spectra compared to the more common A-B-B-A pattern \citep[see the Appendix of][]{kriekMOSFIREDEEPEVOLUTION2015}. 

We perform flat-fielding, slit-tracing, wavelength calibration, sky subtraction, combining, and object detection/extraction using \pypeit$\,$\citep{prochaskaPypeItPythonSpectroscopic2020}. {Our ABC} dither pattern is not supported in the original \pypeit$\,$workflow, so we have manually edited the \texttt{PypeIt Reduction Files} to assign appropriate combination IDs for each frame so that the correct combination of sky frames can be subtracted. We matched each target with the telluric stars by airmass and created a response spectrum for each echellette order by comparing the telluric star spectra with a standard A0V spectrum \citep{picklesStellarSpectralFlux1998}. We used these response spectra to calibrate the science spectra of all targets. The reduced spectra of all orders are combined with \pypeit. {Finally, we perform an absolute flux calibration, also accounting for slit losses, by scaling the NIRES spectra to the WISE photometry.}

\subsection{Emission-line and SFR Measurements}

\begin{figure*}[ht!]
\centering
\includegraphics[width=0.88\textwidth]{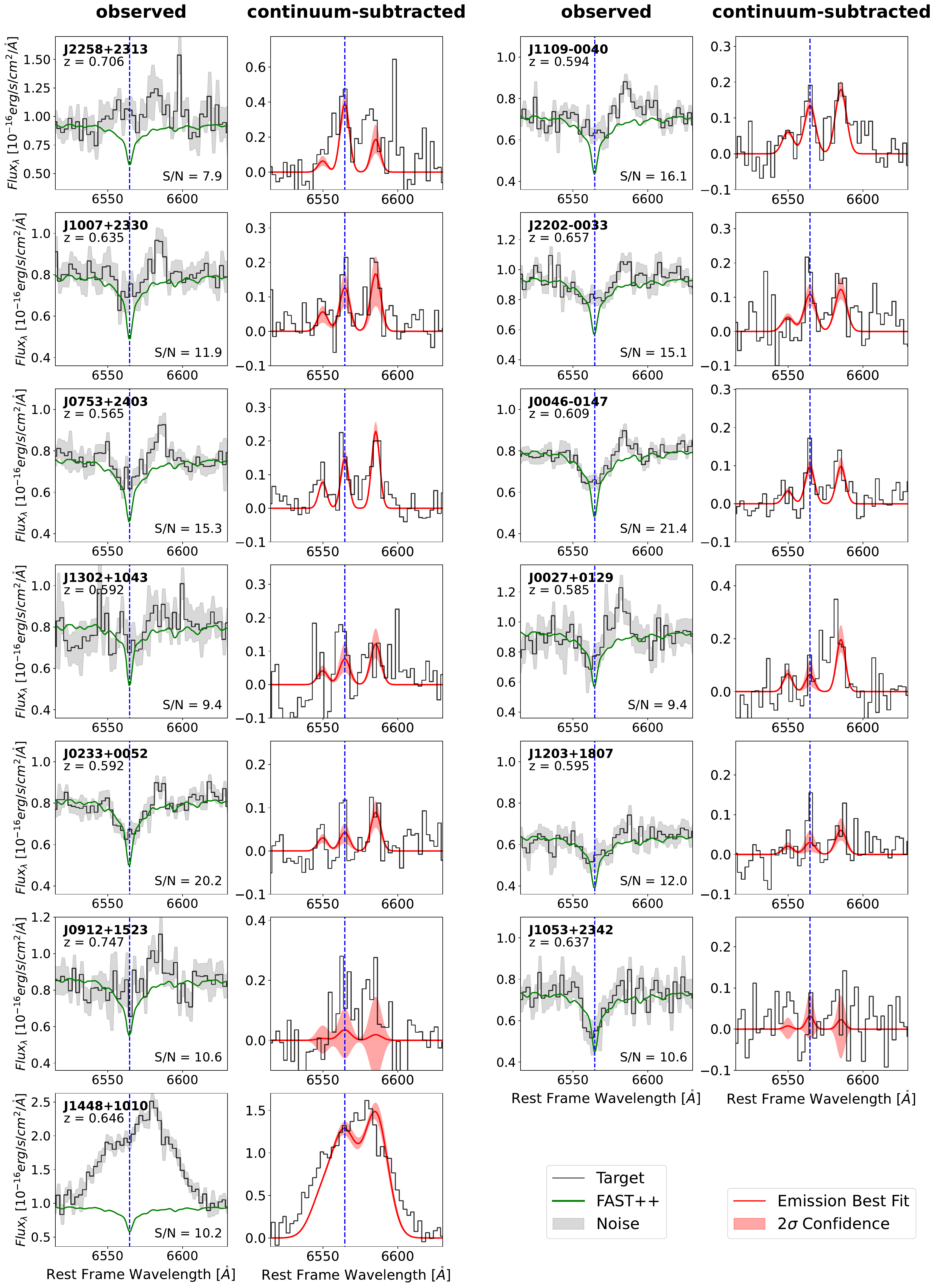}
\caption{Keck/NIRES spectra of all \squiggle PSBs in the $\mathrm{H\alpha}$ region. The panels in the odd-numbered columns display the \texttt{FAST++} modeled continuum in green and the spectra in black; the grey-shaded area represents the noise. The target's SDSS ID and redshift are labeled in the corner. %The grey-shaded areas represent the part of the spectrum used to scale with the continuum, 
{The vertical dashed blue lines mark the $\mathrm{H\alpha}$ wavelength in vacuum (6564.6 $\AA$).} Each panel in the even-numbered columns illustrates the target's emission spectrum after continuum subtraction (in black) corresponding to the left-adjunct panel. The best-fit Gaussian profiles (\ntwo$\mathrm{\lambda6549}$, $\mathrm{H\alpha}$, \ntwo$\mathrm{\lambda6585}$ from left to right) are plotted in red, with the red-shaded areas denoting the 2$\sigma$ confidence intervals. {The median per-pixel S/N over this wavelength range is listed in the lower right corner of each spectrum panel.}}
\label{fig: nires spectra}
\end{figure*}

Figure \ref{fig: nires spectra} displays the reduced and flux-calibrated spectra of the $\mathrm{H\alpha}$ region ($6410{\rm \AA} < \lambda < 6700{\rm \AA}$). We see clear $\mathrm{H\alpha}$ absorption in 11/13 targets, indicative of A-type stars and low SFRs. To derive the SFR, we measure the amount of $\mathrm{H\alpha}$ emission that is infilling this stellar absorption feature by subtracting a stellar continuum model from it. 

{For the continuum we use the best-fit \texttt{FAST++} models, locally scaled to the NIRES spectra (using continuum regions 6410-6530$\AA$ and 6600-6700$\AA$). These fits are preferred over the \texttt{Prospector} fits for this purpose, as they more accurately reproduce the bluer Balmer absorption lines in the SDSS spectra.} Then, we measure the emission-line fluxes from the continuum-subtracted and scaled NIRES spectra, shown in the right panels of Fig~\ref{fig: nires spectra}.

To measure the line fluxes, we fit Gaussian profiles to three emission lines: \ntwo$\mathrm{\lambda6585}$, \ntwo$\mathrm{\lambda6549}$, and $\mathrm{H\alpha}$. We define 4 free parameters for the fitting: the flux of the $\mathrm{H\alpha}$ emission line ${F_\mathrm{H\alpha}}$; the flux of the \ntwo$\mathrm{\lambda6585}$ emission line $F_{\mathrm{[N\, II]\lambda6585}}$; small flexibility $dz$ for the redshift (0 for most of the targets); and the line width in the wavelength space $\sigma_\lambda$. We fix %$F_{[N\, II]\lambda6585}$ also defines the intensity of $\mathrm{[N\, II]\lambda6549}$, since the ratio 
$F_{\mathrm{[N\, II]\lambda6549}}/F_{\mathrm{[N\, II]\lambda6585}} = 0.34$. % is well defined. 
Here $dz$ and $\sigma_\lambda$ are assumed to be identical for the $\mathrm{H\alpha}$ and \ntwo\ lines; this assumption is critical because $\mathrm{H\alpha}$ is typically not detected strongly enough to constrain the line width. We fit for $\sigma_\lambda$ in wavelength space, then convert to the velocity dispersion $\sigma_v$. We allow $\sigma_\lambda$ to vary between the {instrumental} resolution of the Keck/NIRES spectra and the stellar velocity dispersions (with a 2$\sigma$ error) as measured in \citet{suessSQuIGGLEStudyingQuenching2022}. {This broad prior allows for the H$\alpha$ velocity dispersions that may be significantly smaller than the stellar velocity dispersions, as seen in deep spectra of quiescent galaxies at similar redshifts \citep{belli17}.}
%We also take into account the resolution of the Keck/NIRES spectra (defined as $\frac{\Delta\lambda}{\Delta Pixel} /2.355$) as a lower bound, and the velocity dispersion from the \squiggle survey as an upper bound to constrain the Gaussian profiles fitting. The fitted linewidth $\sigma_\lambda$ in wavelength space is converted to the velocity dispersion $\sigma_v$.

{We use Monte Carlo simulations to obtain error bars on the measured $\mathrm{H\alpha}$ and \ntwo\ line fluxes. For each target, we generate 6000 simulated spectra where each data point is drawn from a normal distribution with a mean equal to the observed data point and a standard deviation equal to the observed noise. We re-scale each simulated spectrum to the continuum and re-fit for the line flux, and report the flux measurement as the median of our Monte Carlo samples. We report the 1$\sigma$ confidence interval of the line flux distribution as our error bar. If the flux is not detected at $>2\sigma$, we report 2$\sigma$ upper limits.} Red-shaded areas in the emission panels (right) of Figure \ref{fig: nires spectra} show $2\sigma$ confidence intervals of the fits.

Next, we correct the H$\alpha$ flux for dust attenuation. While $\mathrm{H\beta}$ is covered by the SDSS spectrum for each target, there is no detectable $\mathrm{H\beta}$ emission in the continuum-subtracted spectra for most (12/13) of the targets, which prevents us from using the Balmer decrement to estimate attenuation. Instead, {we dust-correct the fluxes using the stellar attenuation $A_{\mathrm{V,*
}}$ measured by our \texttt{FAST++} modeling. We convert $A_{\mathrm{V,*}}$ to $A_{V, H_{\, II}}$ assuming a \citet{calzettiDustContentOpacity2000} reddening curve and $R_V$ as well as $A_{\mathrm{V,H_{\, II}}}$ / $A_{V,*}=1.86$ following \citealt{priceDIRECTMEASUREMENTSDUST2014}. Because $A_{\mathrm{V}}$ values can be difficult to robustly recover with SED fitting and SFRs can be sensitive to these corrections, we also explored using $A_{\mathrm{V,H_{\, II}}}$ / $A_{V,*}=3$, much higher than the median value from \citet{priceDIRECTMEASUREMENTSDUST2014}, and found that the inferred dust-corrected H$\alpha$ SFRs remain well below the main sequence. We discuss the potential impacts of dust on our measurements in greater detail in Section~\ref{sec: disc}.}

%(converted to $A_{V, H_{\, II}}$ following \citealt{priceDIRECTMEASUREMENTSDUST2014}) from the \texttt{FAST++} modeling, {assuming a \citet{calzettiDustContentOpacity2000} reddening curve and $R_V$.} We note that the \texttt{FAST++} $A_{\mathrm{V,star}}$ could be off by $\sim$a factor of two.

We estimate the SFRs using the \citet{kennicuttStarFormationGalaxies1998} relation, {corrected to a Chabrier IMF following \citet{muzzinWELLSAMPLEDFARINFRAREDSPECTRAL2010}}:
\begin{equation}
\label{eq: sfr}
\mathrm{SFR}(\mathrm{H\alpha})\mathrm{[M_\odot yr^{-1}]} = 4.55 \times 10^{-42}\times L(\mathrm{H\alpha})\mathrm{[erg/s]}
\end{equation}
The measured SFRs and their corresponding errors are listed in Table \ref{tab:all measurements}.

We also measure the line ratios \ntwo$\mathrm{\lambda6585/H\alpha}$ and \othree$\mathrm{\lambda5007/H\beta}$. For the \othree\ and $\mathrm{H\beta}$ emission, we use the continuum-subtracted SDSS spectra since Keck/NIRES does not cover those two lines. We assume the \othree\ and $\mathrm{H\beta}$ lines originate from the same region and thus adopt the same $\sigma_v$ as the $\mathrm{H\alpha}$ and \ntwo\ lines. We calculate the line flux ratios for each of the 6000 Monte Carlo simulated spectra for every target and determine the uncertainties in the fluxes and the ratios. If one of \othree\ or $\mathrm{H\beta}$ is not detected, we determine and report a 2$\sigma$ upper or lower limit on the ratio, respectively. If none of the lines are detected, we cannot place any meaningful constraints on the line ratio.

%This would, in principle, give 6000 line ratio measurements. However, as mentioned above, not all emission lines are detectable in our spectra, especially $\mathrm{H\beta}$. This results in four possible cases: (a) both lines are detected, (b) neither line is detected, (c) the numerator (i.e., $\mathrm{[N\, II]}$ or $\mathrm{[O\, III]}$) is detected (2$\sigma$ lower limit great than zero) and the denominator (i.e., $\mathrm{H\alpha}$ or $\mathrm{H\beta}$) is not, or (d) the denominator is detected while the numerator is not. For case (a), we report the median as the line ratio measurement and 1$\sigma$ confidence interval as error bars, same as the $\mathrm{H\alpha}$ flux measurements. Case (c) means we measure an upper limit, case (d) means we measure a lower limit, and case (b) means that we cannot place any meaningful limits on the line flux ratio. For case (b) or (d), we report the upper limit of the line ratio distribution. For case (c), the denominator's flux is not detected, and we only report the upper limit, but the numerator's flux is detected. After dividing, the denominator's upper limit transforms into a lower limit for the line ratio measurement; we thus report the line ratio distribution's lower limit. 
In the Keck/NIRES spectra, neither $\mathrm{H\alpha}$ nor \ntwo\ is robustly detected for 2 targets (J0912+1523 and J1053+2342); both lines are detected for all the other 11 targets. In the SDSS spectra, both \othree\ and $\mathrm{H\beta}$ are detected only for J0753+2403, only \othree\ is detected for J1448+1010, and neither \othree\ nor $\mathrm{H\beta}$ is detected for all other targets.

\section{Results and Analysis} \label{sec: result}

Figure \ref{fig: nires spectra} shows that clear $\mathrm{H\alpha}$ absorption features are present in the spectra of 11/13 galaxies, {indicating low sSFRs}. As shown in Table \ref{tab:all measurements}, we measure minimal amounts of $\mathrm{H\alpha}$ emission (EW $\leq 4.56\,\mathrm{\AA}$ ) for 12/13 \squiggle PSBs. The galaxies J1053+2342 and J0912+1523 have no measurable $\mathrm{H\alpha}$ emission and we report the $2\sigma$ upper limits on their SFRs. 

The last panel of Figure \ref{fig: nires spectra} shows another noteworthy anomaly, J1448+1010, which is highly likely an AGN given its broad and robust $\mathrm{H\alpha}$ emission and the strong \othree\ emission in its SDSS spectrum {\citep[see detailed discussions in][]{greeneRoleActiveGalactic2020,spilkerStarFormationSuppression2022}}. Consequently, we expect the $\mathrm{H\alpha}$-based SFR of J1448+1010 to be considerably overestimated as its $\mathrm{H\alpha}$ emission most likely arises from an AGN. We thus denote this target with a distinct marker (an empty diamond) in our plots and exclude it from further analyses.

\subsection{Star Formation Rates and $H_2$ Reservoirs} 
\label{subsection: sfr vs H2}

\begin{figure}[h!]
\centering
\includegraphics[width=0.45\textwidth]{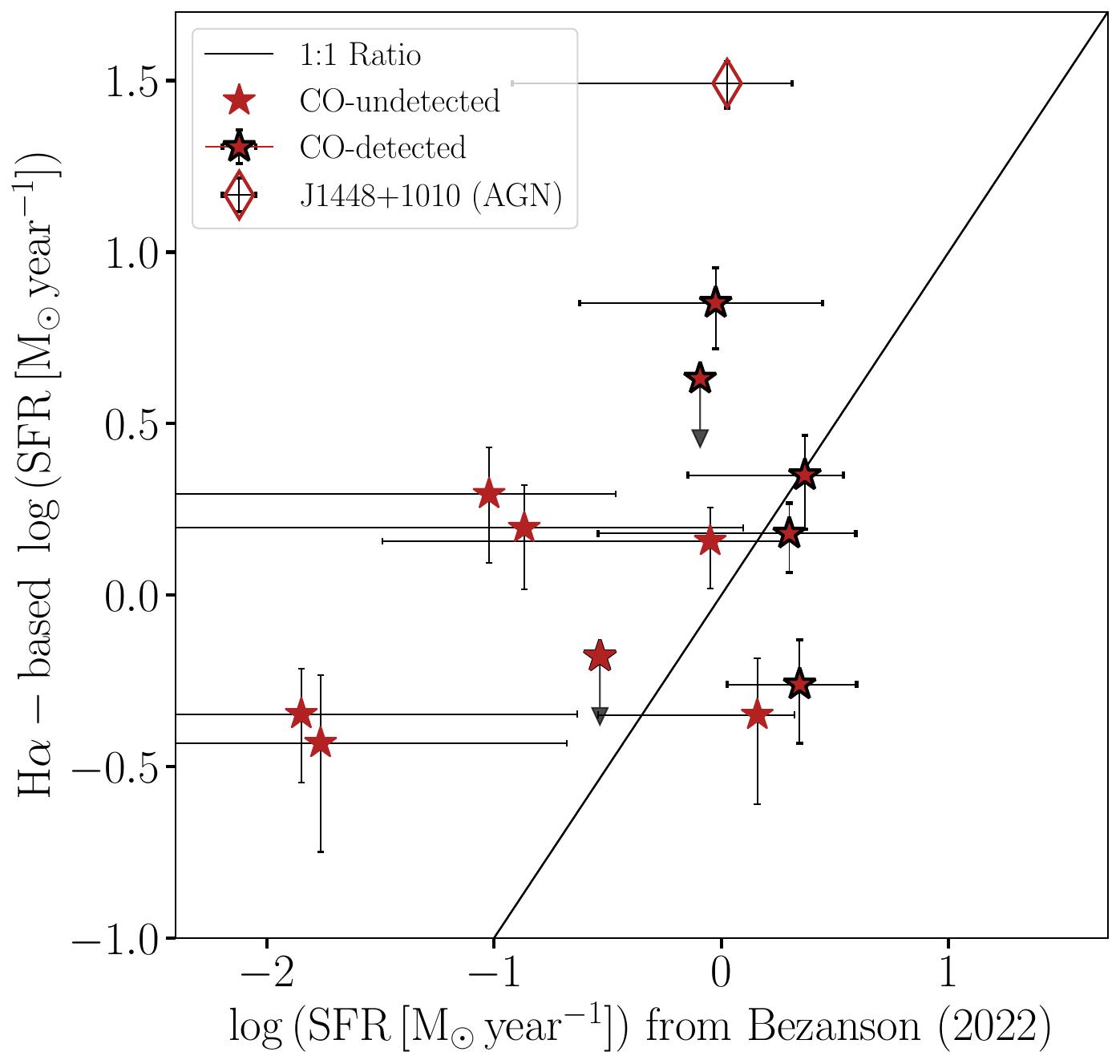}
\caption{Comparison between the SED SFRs measured by \citetalias{bezansonNowYouSee2022}, and the H$\alpha$ SFRs calculated here. The SFR measurements from both studies are generally in good agreement except for J1448+1010, which possesses a broad and intense $\mathrm{H\alpha}$ emission line indicating an AGN.}
\label{fig: sfr vs sfr}
\end{figure}

In Figure \ref{fig: sfr vs sfr}, we compare our Keck/NIRES $\mathrm{H\alpha}$-based SFRs with the SED-based SFRs from \citetalias{bezansonNowYouSee2022}. Overall, the two sets of SFR measurements align well with each other. Again, the exception is J1448+1010, for which the SFR is overestimated {due to AGN contributions, as discussed above}. Most other $\mathrm{H\alpha}$-based SFRs fall within the error range of the corresponding SED-based SFRs. {Given the large error bars on the SED SFRs, the two sets of SFR measurements are statistically consistent. For the 5/12 targets where $\mathrm{H\alpha}$ SFRs are slightly higher than the SED-based SFRs, this offset may be partially due to other ionization mechanisms contributing to the $\mathrm{H\alpha}$ flux (see Sect. \ref{subsection: BPT}), or the fact that the SED-based SFRs are measured over a longer timescale compared to the $\mathrm{H\alpha}$ SFRs.} 

\begin{figure}[h!]
\centering
\includegraphics[width=.45\textwidth]{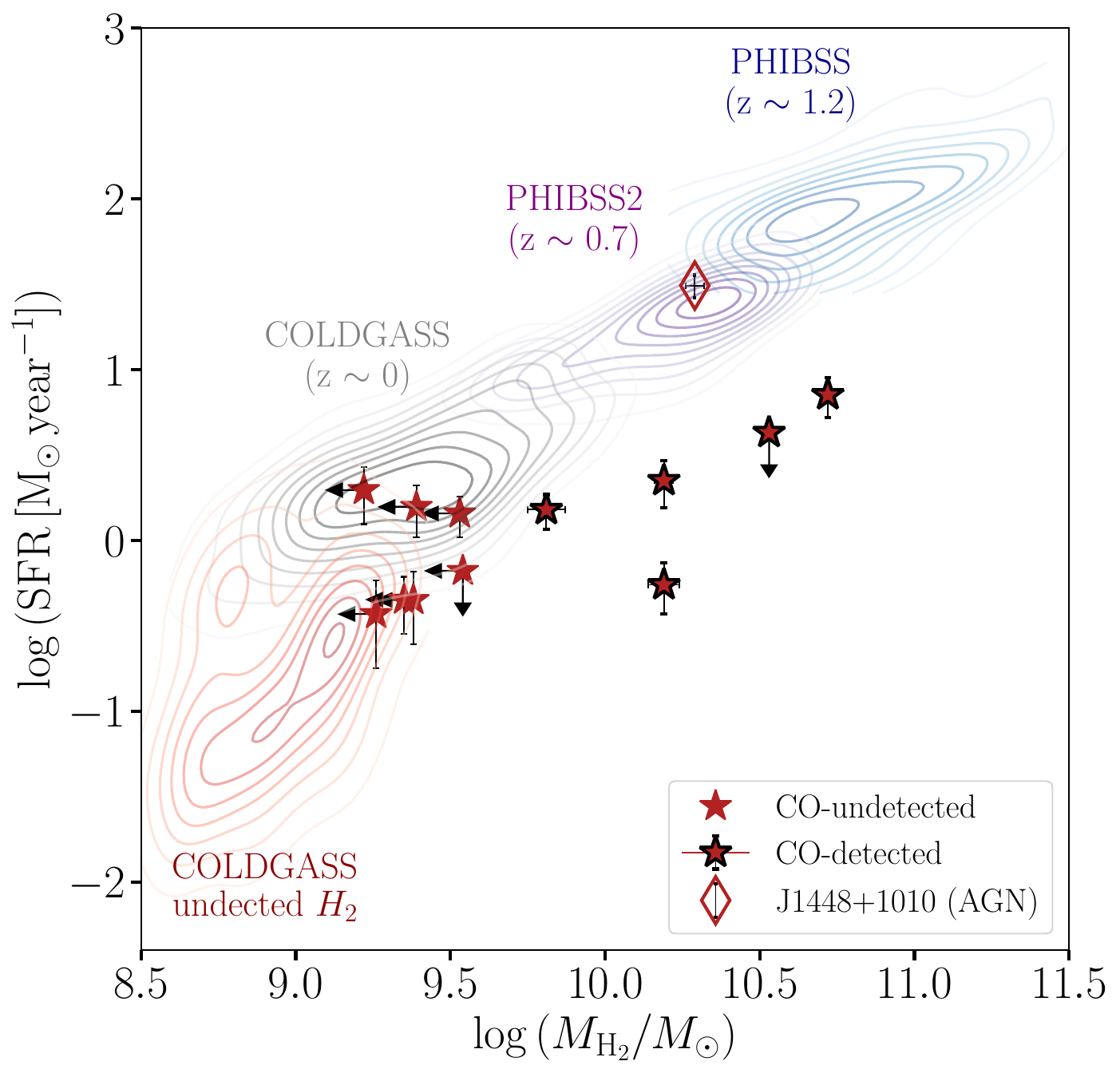}
\caption{The SFR vs. $\mathrm{H_2}$ gas mass for \squiggle PSBs. {The CO-detected targets are plotted as red stars with black edges, and the CO-undetected samples are plotted as red stars.} Comparative samples of massive galaxies from COLDGASS at z $\sim$ 0 \citep{saintongeCOLDGASSIRAM2011, saintongeCOLDGASSIRAM2011a}, {star-forming galaxies at z $\sim$ 0.7 and z$\sim$1.2 from PHIBSS2 and PHIBSS} \citep{tacconiHighMolecularGas2010}, respectively, are depicted as contour plots in the backdrop. Although most galaxies lie close to the $\mathrm{M_{H_2}-SFR}$ relation, several \squiggle post-starburst galaxies possess substantial $\mathrm{H_2}$ reservoirs for their low SFRs. This is most pronounced for the CO(2$-$1)-detected \squiggle galaxies, which are offset by over an order of magnitude in $\mathrm{M_{H_2}}$.}
\label{fig: sfr vs h2}
\end{figure}

In Figure \ref{fig: sfr vs h2}, we compare the scaling relation between the ALMA CO(2-1) $\mathrm{H_2}$ gas mass \citepalias[][assuming $\alpha_\mathrm{CO} = 4.0$]{bezansonNowYouSee2022} and the $\mathrm{H\alpha}$-based SFR measurements of the \squiggle PSBs with that of ``normal" star-forming galaxies at both high and low redshifts. We plot the \squiggle PSBs in red, while the colored contours trace the empirical scaling relation for CO-based $\mathrm{H_2}$ measurements. At lower SFR, massive galaxies at z $\sim$ 0 from the COLDGASS survey \citep{saintongeCOLDGASSIRAM2011, saintongeCOLDGASSIRAM2011a} are depicted in black (for galaxies with detected CO(1-0) lines) and red (upper limits on CO(1-0)) contours. Star-forming galaxies at z = 1.2 from the PHIBSS/PHIBSS2 surveys \citep{tacconiHighMolecularGas2010, genzelPHIBSSMolecularGas2013, freundlichPHIBSS2SurveyDesign2019} are denoted by blue contours. While the ALMA-undetected targets lie close to the $\mathrm{M_{H_2}}$-SFR relation, the ALMA-detected targets have more than an order of magnitude more $\mathrm{M_{H_2}}$ than expected from this relation. Our new $\mathrm{H\alpha}$-based SFRs confirm one of the main findings of \citetalias{bezansonNowYouSee2022}, that large cold gas reservoirs can persist {after quenching despite low levels of ongoing star formation}.%For the CO-undetected targets, the average SFR level is very close to their ``normal" counterparts (COLDGASS z$\sim$0), albeit some are still beneath the principal distribution. 

\subsection{The BPT Diagram} \label{subsection: BPT}

\begin{figure}[t!]
\centering
\includegraphics[width=0.46\textwidth]{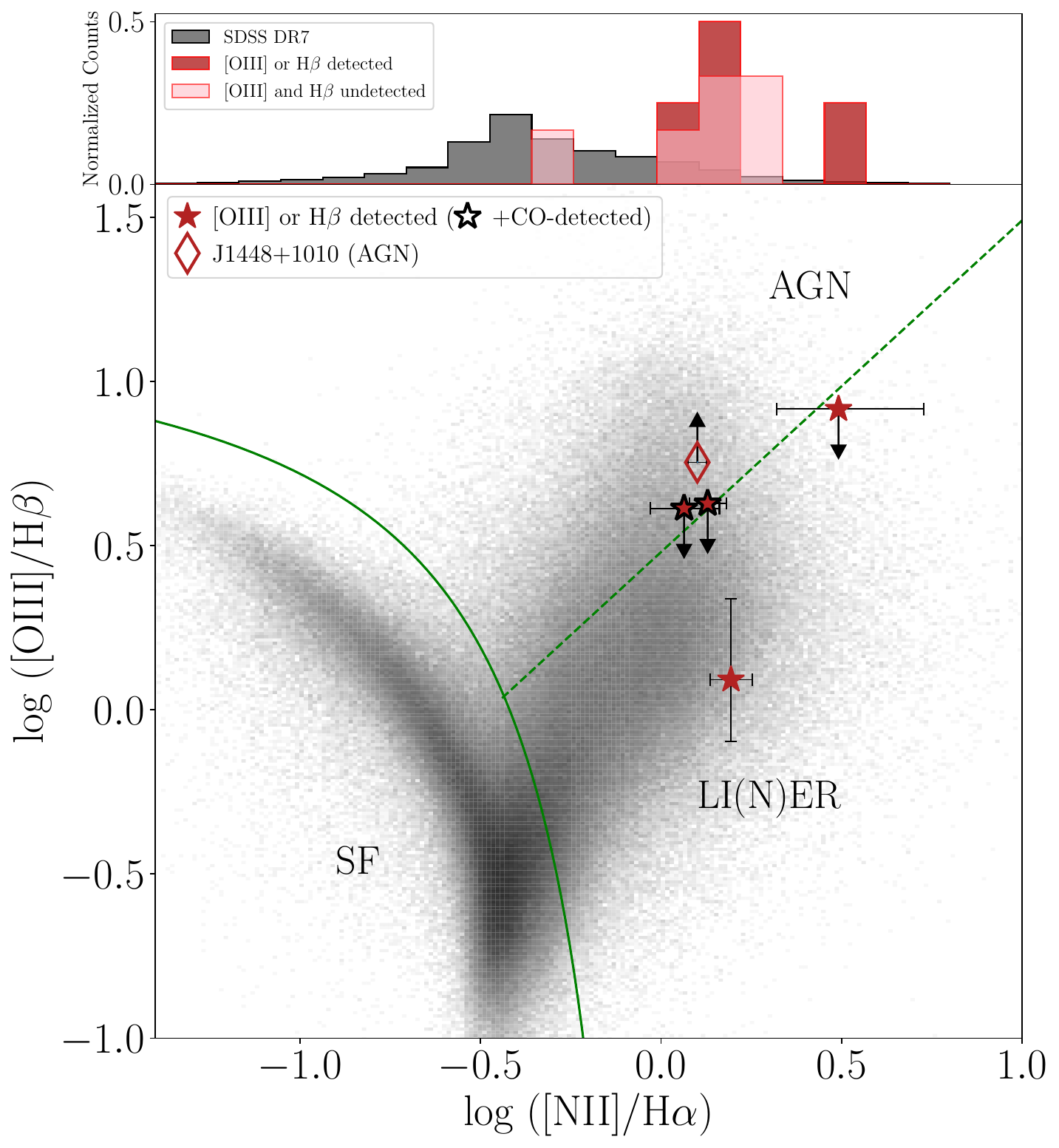}
\caption{{The \squiggle galaxies detected for \ntwo/H$\alpha$ and either H$\beta$ or \othree\, are plotted on the BPT diagram. The green solid curve and the green dashed line show the \citet{kewleyTheoreticalModelingStarburst2001}, and \citet{fernandesAlternativeDiagnosticDiagrams2010} demarcations of star-forming (SF), AGN dominated and LI(N)ER-like galaxies, respectively. The black scatter in the background shows the SDSS DR7 galaxies \citep{abazajianSeventhDataRelease2009}. The histogram on top shows the normalized distribution of the \squiggle galaxies along the $\log$(\ntwo/H$\alpha$) axis, where dark red represents the H$\beta$ or \othree\, detected galaxies, and pink represents the H$\beta$ and \othree\, undetected galaxies. In comparison, the histogram also shows normalized counts of the SDSS galaxies in grey. The two galaxies (J0912+1523 and J1053+2342) that are neither detected in \ntwo/H$\alpha$ nor \othree/H$\beta$ are not shown in this figure.}}
\label{fig: bpt}
\end{figure}

Figure \ref{fig: bpt} shows the \squiggle PSBs on the BPT diagram, with the SDSS DR7 \citep{abazajianSeventhDataRelease2009} galaxies plotted in the background (greyscale) for comparison. {As discussed in Sect. \ref{subsec: obs}, most targets lack a robust \othree/$\mathrm{H\beta}$ line ratio measurement since no $\mathrm{H\beta}$ and/or \othree\, emission is detected. However, we are able to measure \ntwo/H$\alpha$ for 11/13 galaxies. 10/11 of these galaxies fall on the right of the \citet{kewleyTheoreticalModelingStarburst2001} line, as shown in the histogram of Figure \ref{fig: bpt}. Therefore, from this single line ratio alone we can conclude that most galaxies in our sample are likely to be AGN-dominated or LI(N)ER-like \citep[e.g.,][]{yanNatureLINERlikeEmission2012, belfioreSDSSIVMaNGA2016} galaxies. The noteworthy exception J1448+1010 is identified as an AGN by \citet{greeneRoleActiveGalactic2020} both in radio with \texttt{FIRST} and optically via \othree/H$\beta$. }

Thus, the line ratios of our galaxies suggest that ionization mechanisms other than photo-ionization by young massive stars in star-forming regions, like photo-ionization by AGN \citep{halpernLowIonizationActive1983} or by hot evolved (post-AGB) stars \citep{binettePhotoionizationEllipticalGalaxies1994}, or fast radiative shocks \citep{dopitaSpectralSignaturesFast1996}, could be contributing to or dominating the $\mathrm{H\alpha}$ emission. Consequently, the $\mathrm{H\alpha}$ SFRs we calculate in this paper could overestimate the ongoing SFR. While the $\mathrm{H\alpha}$ SFRs we measure for \squiggle PSBs are already low, the BPT diagram indicates that the true SFRs of these galaxies could be even lower.

\section{Discussion and Conclusions} \label{sec: disc}

This paper presents new dust-corrected $\mathrm{H\alpha}$ measurements confirming low ongoing SFRs in gas-rich PSBs at z$\sim$0.6. Our results support the conclusion of \citetalias{bezansonNowYouSee2022} that these galaxies shut down their star formation before their molecular gas reservoirs were fully depleted. Except for J1448+1010, which has strong AGN-contributed $\mathrm{H\alpha}$ flux, {the galaxies in our sample have low $\mathrm{H\alpha}$ SFRs $\sim 0.4-7 \,\, \mathrm{M_\odot yr^{-1}}$, indicating that they have already shut down active star formation. }Therefore, the large molecular gas reservoirs ($\sim 1-5 \times 10^{10} \,\, \mathrm{M_\odot}$) present in $\sim$half of these galaxies {cannot be explained by residual ongoing star formation that was missed by our previous study \citetalias{bezansonNowYouSee2022}}. Moreover, their positioning on the BPT diagram implies that {star formation is not the dominant ionization source for any of the 13 galaxies in this sample,} indicating that their true SFRs may be even lower than reported here.

%One of our goals in this study is to understand the presence of dust-obscured star formation in the \squiggle PSBs. We find no evidence of strong $\mathrm{H\alpha}$ emission in most of the Keck/NIRES spectra; the one exception is J1448+1010, discussed in detail in \citet{spilkerStarFormationSuppression2022}, where the broad $\mathrm{H\alpha}$ is due to the presence of an AGN. The galaxies J0912+1523 and J2258+2313 show weak $\mathrm{H\alpha}$ emission, 
%Most (10/13) galaxies in our sample show net $\mathrm{H\alpha}$ absorption, with only minimal infilling due to ongoing star formation. 
 %Furthermore, we find no evidence of strong $\mathrm{H\alpha}$ emission in most of the Keck/NIRES spectra (excluding J1448+1010, for which we attribute the emission to AGN activity). The galaxies J0912+1523 and J2258+2313 displayed more pronounced $\mathrm{H\alpha}$ emission, yet, as illustrated in Table \ref{tab:all measurements} and Figure \ref{fig: sfr vs h2}, their SFR measurements remain low considering their substantial $\mathrm{H_2}$ masses. However, we recognize the potential oversight of undetected obscured star formation. We didn't use the Balmer decrements to estimate the attenuation since the $\mathrm{H\beta}$ emission is barely detectable in the SDSS spectra, which may indicate more dust-obscured SFRs if used.

{Based on} their \ntwo/$\mathrm{H\alpha}$ ratios, all the \squiggle PSBs in this sample could potentially harbor some level of AGN activity. This finding is consistent with the optically identified AGN occurrence rates reported by \citet{greeneRoleActiveGalactic2020} via \othree/H$\beta$ ratios for the same parent sample: they find that the youngest \squiggle galaxies are $\sim10$ times more likely to harbor an optical AGN {than an older control sample}, indicating {that AGN activity may have played a role in quenching } \citep[as theoretically predicted by, e.g.][]{hopkinsCosmologicalFrameworkCoEvolution2008}. However, {we note that the observed high \ntwo/$\mathrm{H\alpha}$ ratios in our NIRES spectra could also be explained by contributions from {hot evolved stars} \citep[seen in low-z quiescent galaxies, e.g.,][]{yanNatureLINERlikeEmission2012} and/or shocks \citep[e.g.,][]{alataloShockedPOststarburstGalaxy2016}. Shocks may be expected in this sample given that many \squiggle PSBs, especially the youngest galaxies which are the most likely to host molecular gas reservoirs, show clear signs of recent mergers \citep{verricoMergerSignaturesAre2023}. \citet{wuStarsGasStar2023} find that companion galaxies and low surface brightness features are common around a sample of five $z\sim0.7$ PSBs from the LEGA-C survey, four of which are gas-rich, supporting the idea that mergers are a key player in quenching these galaxies. These mergers may have added more gas {\it after} the primary quenching event in these PSBs \citep{woodrumMolecularGasReservoirs2022}.
}

{The $\mathrm{H\alpha}$ SFRs reported in this work place the CO-detected \squiggle galaxies below the star-forming main sequence and significantly offset from the Kennicutt-Schmidt relation, with larger gas reservoirs than expected given their low instantaneous SFRs. Similar conclusions have also been reported in several samples of PSBs in the local universe \citep[e.g.,][]{frenchDiscoveryLargeMolecular2015, rowlandsEvolutionColdInterstellar2015}. The question remains why this molecular gas persists after star formation shuts down. Here, we discuss several possibilities. 

The first option is that the gas-rich quiescent PSBs in our sample only appear offset from the Kennicutt-Schmidt relation due to observational biases. Despite our best efforts to account for dust, it is possible that completely optically-thick dust could mask ongoing star formation. In this work, we corrected our $\mathrm{H\alpha}$ fluxes for dust using the best-fit A$_V$ from stellar population synthesis fitting; because no $\mathrm{H\beta}$ infilling is detected for 12/13 of our galaxies, Balmer decrement corrections are unfortunately impossible on an individual galaxy basis. These SED-fitting A$_V$ measurements could potentially underestimate the true dust attenuation in these galaxies. While ALMA 2mm non-detections from \citet{bezansonExtremelyLowMolecular2019} imply obscured SFR$\lesssim 50 \,\, \mathrm{M_\odot /yr}$, lower levels of $\sim10\mathrm{M_\odot /yr}$ would be sufficient to place our gas-rich targets on the Kennicutt-Schmidt relation. { Given our measured H$\alpha$ fluxes, reaching these SFRs would require an average $A_{\mathrm{V, H_{II}}}$ of 4.4~mag -- higher than the typical $A_{\mathrm{V, H_{II}}}$ values measured in local PSBs by \citet{yesufFilthyRichDiverse2020}, which peak at $A_{\mathrm{V, H_{II}}}\sim$1~mag but show a wide range of $0 \lesssim A_{\mathrm{V, H_{II}}}\lesssim 3$~mag.} \citet{poggiantiOpticalSpectralSignatures2000} and \citet{baronStarFormationMolecular2023} also show that some galaxies with low optical SFRs can have significantly higher FIR SFRs. However, the galaxies in these samples with high FIR SFRs uniformly show H$\alpha$ in {\it emission} { even when other Balmer lines are seen in absorption. This H$\alpha$ emission is distinct from the strong absorption we find in our sample}, implying that these results are not directly applicable to our galaxies, where any residual star formation would have to be obscured by completely optically-thick dust in order to produce strong H$\alpha$ absorption. Furthermore, IR SFRs are a biased tracer for post-starburst and quiescent galaxies due to heating by old and/or evolved stars \citep[e.g.,][]{utomo14,wild24}. While we cannot completely rule out the possibility of dust-obscured star formation, given our measurements in this work, we believe this scenario to be unlikely.

Another potential bias is converting CO(2-1) to molecular gas. { This conversion relies on both $r_{21}$ and $\alpha_{\rm{CO}}$. We have conservatively assumed thermalized CO, $r_{21} = 1.0$, the highest possible value: departures from this assumption would only increase the inferred molecular gas masses. We have assumed a Milky Way-like $\alpha_{\rm{CO}}=4.0$ \citep[e.g.][and references therein]{bolattoCOH2ConversionFactor2013}. As discussed in \citet{bezansonNowYouSee2022}, lower values of $\alpha_{\rm{CO}}$ are typical of dusty, highly star-forming, and turbulent galaxies. These conditions increase the CO excitation temperature and lower the line opacity, allowing more CO luminosity to escape per unit mass \citep[e.g.,][]{narayananGeneralModelCOH22012}. As our galaxies do not meet these conditions, we presently have no reason to believe that a lower value of $\alpha_{\rm{CO}}$ is appropriate for our sample. That being said, it is possible that the CO-rich galaxies in our sample may lack the very dense molecular gas required to form stars,} as hinted at by local observations of PSBs which lack denser gas tracers HCN and HCO+ \citep[e.g.,][]{frenchClockingEvolutionPoststarburst2018}. These dense gas tracers are practically inaccessible at our $z\sim0.6$ redshift range, but future studies of higher-J CO transitions may provide additional insights into the density and and state of the interstellar medium of these galaxies. 

The second option is that the gas-rich quiescent PSBs in our sample {\it are} truly offset from the Kennicutt-Schmidt relation, as implied by our measurements in this work. This reinforces the conclusion that galaxies can quench while retaining substantial molecular gas reservoirs, and implies that the process that quenched these galaxies did not remove the molecular gas. One such quenching mechanism is ``morphological quenching," where dynamical support against molecular gas collapse can result from turbulent gas pressure \citep{martigMORPHOLOGICALQUENCHINGSTAR2009}. \citet{gensiorHeartDarknessInfluence2020} found that compact, spheroidal structures can drive this turbulent pressure and result in global SFRs suppressed by factors of $\sim5$ and central SFRs suppressed by up to two orders of magnitude --- not dissimilar from the factor of $\sim10$ offset we see between our gas-rich PSBs and the Kennicutt-Schmidt relation in Figure~\ref{fig: sfr vs h2}.} Morphological quenching could be especially relevant for our sample given that the sizes of PSBs are often compact, even compared to older quiescent galaxies \citep[e.g.,][]{whitaker12,yanoRelationGalaxyStructure2016,wuColorsSizesRecently2020,suessDissectingSizeMassS1Mass2021,settonCompactStructuresMassive2022}. Our observations of high [N II]/H$\alpha$ ratios suggest that {AGN may have also played a role in the quenching process of the \squiggle PSBs}. However, existing quenching models { often} assume that AGN feedback removes or heats the molecular gas \citep[e.g.,][]{hopkinsUnifiedMergerdrivenModel2006} rather than preserving it, { unlike our observations of gas-rich quiescent PSBs.} { That being said, identifying exact analogs to \squiggle-like galaxies in simulations is challenging both because it involves finding very rare objects in limited simulation volumes and due to uncertain gas post-processing. Several studies focusing on quenching galaxies in simulations have suggested that these galaxies may be able to retain molecular gas during the transition phase \citep{davies21}, perhaps because the gas is too kinematically disturbed to collapse into stars \citep{davis19}.}%Moreover, as most of these processes can only decrease the SFR by a relatively small factor, it is unlikely that any models can adequately address the observed significant disparity between this \squiggle sample and the star-forming main sequence. %One simulation shows how relativistic jets can trigger shocks that reduce SFRs by a factor of $\sim$2 \citep[e.g.,][]{mandalImpactRelativisticJets2021}. However, there are also instances where jets cannot effectively quench star formation \citep[e.g.,][]{suWhichAGNJets2021}. Moreover, as most of these processes can only decrease the SFR by a relatively small factor, it is unlikely that any feedback models can adequately address the observed significant disparity between this \squiggle sample and the star-forming main sequence. 

%As we cannot correct the attenuation using the Balmer decrement, we still note the possibility of dust-obscured star formation. Nonetheless, based on non-detections of 2mm continuum in the ALMA data, \citetalias{bezansonNowYouSee2022} ruled out large amounts of obscured star formation ($\lesssim 50 \,\, \mathrm{M_\odot /yr}$), but with large uncertainties. We also note that the $\mathrm{H_2}$-rich \squiggle galaxies may have only temporarily halted their star formation. 

In future studies, deeper observations of the ALMA-undetected targets could {determine whether the transition from gas-rich to gas-poor PSBs happens smoothly}. {Additionally, spatially resolved CO(2-1) maps or observations of higher-J CO transitions might provide insights into the distribution, kinematics, and density of the gas}, potentially shedding light on its resistance to collapse.

%Since we do not reliably detect any H$\beta$ emission infilling the stellar absorption feature (except for J1448+1010), we cannot calculate Balmer decrement-corrected SFRs; instead, we use the attenuation estimated by the \texttt{FAST++} modeling to measure dust-corrected $\mathrm{H\alpha}$ SFRs, which is in general with less constraints. Thus, we still note the possibility of dust-obscured star formation. However, we find no evidence of significant dust-obscured SFRs, as most (11/13) targets show net $\mathrm{H\alpha}$ absorption. This result is consistent with \citetalias{bezansonNowYouSee2022}, who reported upper limits of SFR$\lesssim 50 \,\, M_\odot yr^{-1}$ from the non-detection of continuum emission in the ALMA datacubes.

\begin{acknowledgments}

{We thank the anonymous referee for a detailed report which helped improve this manuscript. }
P.Z. acknowledges the valuable help from Chloe Chen, Daming Yang, David Blanquez Sese, and the \pypeit$\,$development team, without which the data reduction cannot be done, and the help from Yilun Ma for producing the BPT diagram. D.S. gratefully acknowledges the support provided by The Brinson Foundation through a Brinson Prize Fellowship grant for this work. We gratefully acknowledge support from NSF-AAG \#1907697, \#1907723, and \#1908137. {This material is based upon work supported by the National Science Foundation under Grant No. 2407954 \& 2407955.} The infrared spectral data presented in this paper were obtained at the W.M. Keck Observatory, which is operated as a scientific partnership among the California Institute of Technology, the University of California, and the National Aeronautics and Space Administration. The Observatory was made possible by the generous financial support of the W.M. Keck Foundation. The authors wish to recognize and acknowledge the significant cultural role and reverence that the summit of Mauna Kea has always had within the indigenous Hawaiian community. We are most fortunate to have the opportunity to conduct observations from this mountain. {This paper uses the following ALMA data: ADS/JAO.ALMA \#2016.1.01126.S and ADS/JAO.ALMA \#2017.1.01109.S. ALMA is a partnership of ESO (representing its member states), NSF (USA), and NINS (Japan), together with NRC (Canada), MOST and ASIAA (Taiwan), and KASI (Republic of Korea), in cooperation with the Republic of Chile. The Joint ALMA Observatory is operated by ESO, AUI/NRAO and NAOJ.} 

\end{acknowledgments}

%% To help institutions obtain information on the effectiveness of their 
%% telescopes the AAS Journals has created a group of keywords for telescope 
%% facilities.
%
%% Following the acknowledgments section, use the following syntax and the
%% \facility{} or \facilities{} macros to list the keywords of facilities used 
%% in the research for the paper.  Each keyword is check against the master 
%% list during copy editing.  Individual instruments can be provided in 
%% parentheses, after the keyword, but they are not verified.

%\vspace{5mm}
%\facilities{}

%% Similar to \facility{}, there is the optional \software command to allow 
%% authors a place to specify which programs were used during the creation of 
%% the manuscript. Authors should list each code and include either a
%% citation or url to the code inside ()s when available.

%\software{}

\bibliography{main_revised_v2.bbl}{}
\bibliographystyle{aasjournal}

%% This command is needed to show the entire author+affiliation list when
%% the collaboration and author truncation commands are used.  It has to
%% go at the end of the manuscript.
%\allauthors

%% Include this line if you are using the \added, \replaced, \deleted
%% commands to see a summary list of all changes at the end of the article.
%\listofchanges

\end{CJK*}
\end{document}